\documentclass[conference]{IEEEtran}
\usepackage{latexsym}
\usepackage{graphicx}
\usepackage{amsfonts,amssymb,amsmath}
\usepackage{ctable} 
\usepackage{mathtools, cuted}
\usepackage{hyperref}
\usepackage[ruled,linesnumbered,noend]{algorithm2e}
\usepackage[T1]{fontenc}
\usepackage{cite}
\usepackage{subcaption}
\usepackage{comment}
\usepackage{amsthm}
\usepackage{overpic}
\usepackage{steinmetz}
\usepackage{array}
\usepackage{url}
\usepackage{color}
\IEEEoverridecommandlockouts
\usepackage{algorithmicx}
\usepackage{algcompatible}
\usepackage{xcolor}
\usepackage{soul}
\theoremstyle{plain}

\newtheorem{remark}{Remark}

\newcommand{\argmax}[1]{{\underset{{#1}}{\mathrm{arg\,max}}}}

\newcommand{\vect}[1]{\mathbf{#1}}

\newcommand{\bl}[1]{\boldsymbol{#1}}

\def\diag{\mathrm{diag}}

\def\T{\mathrm{T}}
\def\H{\mathrm{H}}

\def\mod{\mathrm{mod}}

\def\T{\mathrm{T}}
\def\H{\mathrm{H}}

\begin{document}

\title{An Efficient Modified MUSIC Algorithm for RIS-Assisted Near-Field Localization   }

\author{Parisa Ramezani, Alva Kosasih, and Emil Bj\"{o}rnson\\
\IEEEauthorblockA{\textit{ Department of Computer Science, KTH Royal Institute of Technology, Stockholm, Sweden} \\  Email: \{parram, kosasih, emilbjo\}@kth.se}%
\thanks{This work was supported by the FFL18-0277 grant and SUCCESS project (FUS21-0026), funded by the Swedish Foundation for Strategic Research.}}

\maketitle

\begin{abstract}
   In this paper, we consider a single-anchor localization system assisted by a reconfigurable intelligent surface (RIS), where the objective is to localize multiple user equipments (UEs) placed in the radiative near-field region of the RIS by estimating their azimuth angle-of-arrival (AoA), elevation AoA, and distance to the surface. The three-dimensional (3D) locations can be accurately estimated via the conventional  MUltiple SIgnal Classification (MUSIC) algorithm, albeit at the expense of tremendous complexity due to the 3D grid search.  In this paper, capitalizing on the symmetric structure of the RIS, we propose a novel modified MUSIC algorithm that can efficiently decouple the AoA and distance estimation problems and drastically reduce the complexity compared to the standard 3D MUSIC algorithm. Additionally, we introduce a spatial smoothing method by partitioning the RIS into overlapping sub-RISs to address the rank-deficiency issue in the signal covariance matrix. We corroborate the effectiveness of the proposed algorithm via numerical simulations and show that it can achieve the same performance as 3D MUSIC but with much lower complexity. 
\end{abstract}
\begin{IEEEkeywords}
Reconfigurable intelligent surface, near-field localization, MUSIC algorithm.
\end{IEEEkeywords}

\section{Introduction}
Recently, reconfigurable intelligent surface (RIS) has sprung up as a revolutionary technology that can control signal propagation by dynamically manipulating incident electromagnetic waves before reflecting them. RIS has been proved effective for enhancing communication performance, particularly in scenarios where the direct link between the transmitter-receiver pair is severely impaired \cite{Emil2022SPM}. Besides its great potential for enhancing communication performance, RIS is also envisioned to contribute to wireless localization thanks to its typically large size and reconfiguration capability \cite{Henk2020VTM}.

Wireless localization has been a fundamental problem in array signal processing for decades with diverse applications in sonar, radar, seismic exploration, and wireless communications \cite{Wax1983,Malioutov}. With the new envisaged location-based services and the need for accurate location information to fulfill the ambitious communication goals, localization is of increasing importance for the upcoming generations of wireless networks \cite{xiao2022overview}.  
Over the past decades, various methods have been developed to address this problem, among which subspace-based methods such as MUltiple SIgnal Classification (MUSIC) and Estimation of Signal Parameters via Rotational Invariance Techniques (ESPRIT) and their variations have received considerable attention due to their high accuracy \cite{stoica2005spectral}.  
The principal idea underlying these methods is to separate the signal and noise subspaces of the signals received at an antenna array and exploit the orthogonality between the two subspaces to estimate the angle-of-arrival (AoA) of different user equipments (UEs). 

When the UE is located in the radiative near-field region of the antenna array, the spherical wavefronts create circular phase variations over the array. These variations are characterized by both the AoA and the distance between the UE and the array, necessitating the joint estimation of AoA and distance for locating the UE. Several algorithms have been devised for near-field localization building upon the conventional MUSIC and ESPRIT localization methods. One of the earliest works on the topic is \cite{Huang1991}, where the classical MUSIC algorithm, which finds the AoAs by performing a grid search over all possible angles, is extended to two-dimensional (2D) MUSIC. In particular, a grid search over two dimensions, i.e., azimuth AoA and distance, is carried out for simultaneous estimation of both parameters. Utilizing the symmetric structure of an antenna array, some algorithms have been later developed that, by decoupling the AoA and distance estimation, reduce the complexity of joint AoA and distance estimation \cite{He2012efficient},\cite{Zhi2007nearfield}. In \cite{He2012efficient}, the AoAs are first estimated by finding the peaks of a MUSIC-like spectrum that is formed by the anti-diagonal entries of the covariance matrix. Thereafter, the corresponding distances are estimated using the standard MUSIC algorithm. The authors in \cite{Zhi2007nearfield} divide the array into two symmetric subarrays and apply a generalized ESPRIT algorithm \cite{Gao2005generalized} to first find the AoAs and then the respective distances. 

RIS can be of specific importance for aiding single-anchor localization with a blocked direct channel where the  links between the base station (BS) and the UEs are obstructed, and the signals sent by the UEs have to travel through the RIS to reach the BS. 
Similar to traditional near-field localization, joint AoA and distance estimation is required in RIS-assisted localization if the UEs are located in the near-field region of the surface, a scenario which is highly probable due to the large aperture of the RIS and the evolution towards higher frequencies \cite{Ramezani2023Exploiting}. This implies simultaneous azimuth AoA, elevation AoA, and distance estimation since an RIS is generally a uniform planar array (UPA). The standard MUSIC algorithm can be employed in this case to jointly estimate all three parameters of the UEs' locations by performing a grid search in the three-dimensional (3D) domain. 
However, this approach has an exceedingly high computational complexity. Therefore, developing efficient near-field localization algorithms for RIS-assisted systems is essential. The reference \cite{Pan2023RIS} is a pioneering work on the design of subspace-based near-field localization algorithms for RIS-assisted systems, where the authors decouple the estimation of the three parameters and separately estimate azimuth AoA, elevation AoA, and distance. Nevertheless, the algorithm proposed in \cite{Pan2023RIS} involves several matrix inversions, the complexity of which rapidly grows with the number of RIS elements and UEs. This makes the applicability of the developed algorithm to practical scenarios questionable. 

Inspired by the work in \cite{He2012efficient}, we propose a new efficient localization algorithm in this paper for estimating the 3D location of multiple near-field UEs in an RIS-assisted system. We first apply a least squares (LS) method to obtain the signal incident on the RIS. Based on the structure of the covariance matrix of the incident signal, we decouple the AoA and distance estimation problems. To deal with the rank-deficiency of the signal covariance matrix, we propose a spatial smoothing method by partitioning the RIS into multiple overlapping sub-RISs. 
As the sample covariance matrix of the incident signal on the sub-RISs has full column rank, we can form a MUSIC-like spectrum and find the azimuth and elevation AoAs via a 2D grid search. The corresponding ranges can be obtained by substituting the estimated AoAs into the array response expression and applying the standard MUSIC algorithm. Our numerical results demonstrate that the proposed algorithm has a comparable performance to the 3D MUSIC algorithm, while having a significantly lower complexity. 

\vspace{-5mm}

\section{System Model}
\vspace{-3mm}
We consider an RIS-assisted  system where the BS is equipped with $M$ antennas, the RIS has $N$ elements in the form of a UPA with $N_{\mathrm{H}}$ elements in each horizontal row and $N_{\mathrm{V}}$ elements in each vertical column such that $N = N_{\mathrm{H}} N_{\mathrm{V}}$, and both $N_{\mathrm{H}}$ and $N_{\mathrm{V}}$ are assumed to be odd numbers. The RIS is located in the YZ plane and the element at the center of the RIS is assumed to be the reference element. There are $K$ single-antenna UEs that need to be localized and the direct links between the UEs and the BS are assumed to be blocked. As the location of the BS and the RIS are fixed, the channel between them is assumed to be static and known, while the $K$ UEs are located at unknown locations in the radiative near-field region of the RIS.

Let $d_{\mathrm{H}}$ and $d_{\mathrm{V}}$ represent the inter-element spacing between adjacent horizontal and vertical RIS elements, respectively. Further, let $n_{\mathrm{H}}$ and $n_{\mathrm{V}}$ denote the horizontal and vertical indices of an element, respectively, with $n_{\mathrm{H}} = - \frac{N_{\mathrm{H}} - 1}{2},\ldots,\frac{N_{\mathrm{H}} - 1}{2}$ and $n_{\mathrm{V}} = - \frac{N_{\mathrm{V}} - 1}{2},\ldots,\frac{N_{\mathrm{V}} - 1}{2}$. Assume that the $k$th UE is located at $r_k (\cos(\varphi_k)\cos(\theta_k),\sin(\varphi_k)\cos(\theta_k),\sin(\theta_k))$, where $r_k$ is the distance from the UE to the reference RIS element, and $\varphi_k$ and $\theta_k$ respectively represent the azimuth and elevation AoA from the UE to the RIS. The distance between the $k$th UE and the RIS element with the horizontal index $n_{\mathrm{H}}$ and vertical index $n_{\mathrm{V}}$ is obtained as 
\begin{align}
\label{eq:distance_exact}
 &r_k^{(n_{\mathrm{H}},n_{\mathrm{V}})} =
\Big(\left(r_k\cos(\varphi_k)\cos(\theta_k)\right)^2+ \nonumber \\
&\left(r_k \sin(\varphi_k)\cos(\theta_k) - n_{\mathrm{H}} d_{\mathrm{H}}\right)^2 + \left(r_k \sin(\theta_k) - n_{\mathrm{V}} d_{\mathrm{V}}\right)^2 \Big)^{1/2}    \nonumber \\
&  =  r_k \bigg( 1  - \frac{2n_{\mathrm{H}} d_{\mathrm{H}} \sin(\varphi_k) \cos(\theta_k)}{r_k} - \frac{2n_{\mathrm{V}} d_{\mathrm{V}} \sin(\theta_k)}{r_k} \nonumber \\ & + \frac{n_{\mathrm{H}}^2 d_{\mathrm{H}}^2 + n_{\mathrm{V}}^2 d_{\mathrm{V}}^2}{r_k^2}\bigg)^{1/2} .
\end{align}

In time-slot $t$, the $k$th UE transmits the signal $s_k(t)$ which is a random zero-mean signal that might contain data, is unknown to the BS, and satisfies $\mathbb{E}\{s_k(t)s_k^*(t)\} = 1$. Furthermore, the signals of the $K$ UEs are assumed to be independent. The received signal at the BS at time-slot $t$ is given by 
\begin{equation}
\label{eq:received_signal}
   \vect{y}(t) = \sum_{k=1}^K \sqrt{p_k} \vect{H}\bl{\Phi} \vect{g}_k s_k(t) + \vect{w}(t), 
\end{equation}where $p_k$ is the transmit power of the $k$th UE,
$\vect{H} \in \mathbb{C}^{M\times N}$ is the known channel matrix between the RIS and the BS, 
 $\bl{\Phi} = \diag(e^{j\phi_1},\ldots,e^{j\phi_N})$ is the RIS reflection matrix with $\phi_n$ being the phase shift applied by the $n$th RIS element to the incident signal, and $\vect{g}_k = \sqrt{\eta_k} \vect{a}(\varphi_k,\theta_k,r_k)$ is the channel between the $k$th UE and the RIS with $\vect{a}(\cdot)$ denoting the RIS's near-field array response vector and $\eta_k$ being the path-loss. 
Moreover, $\vect{w}(t)= [w_1(t),\ldots,w_M(t)]^\T$ is the additive independent complex Gaussian noise with variance $\sigma^2$ across all BS antennas. 

We number RIS elements from the bottom left, row by row. Therefore, the first RIS element is the one with the horizontal index $n_{\mathrm{H}} = - \frac{N_{\mathrm{H}} - 1}{2}$ and the vertical index $n_{\mathrm{V}} = - \frac{N_{\mathrm{V}} - 1}{2}$. Similarly, the last RIS element is the one with $n_{\mathrm{H}} = \frac{N_{\mathrm{H}} - 1}{2}$ and $n_V =  \frac{N_{\mathrm{V}} - 1}{2}$. The RIS array response vector is given by 
\begin{align}
\label{eq:array_response_vector}
   &\vect{a}(\varphi_k,\theta_k,r_k) = \left[e^{j\frac{2\pi}{\lambda}\left(r_k - r_k^{\left(- \frac{N_{\mathrm{H}} - 1}{2},- \frac{N_{\mathrm{V}} - 1}{2}\right) }\right) },\ldots, \right. \nonumber \\  &\left. e^{j\frac{2\pi}{\lambda}\left(r_k - r_k^{\left( \frac{N_{\mathrm{H}} - 1}{2},\frac{N_{\mathrm{V}} - 1}{2}\right) }\right) }\right]^\T,
\end{align}
where $\lambda$ denotes the wavelength. Using Fresnel approximation, the distance expression in \eqref{eq:distance_exact} can be approximated as 
\begin{align}
\label{eq:distance_approx}
&r_k^{(n_{\mathrm{H}},n_{\mathrm{V}})} \approx \nonumber \\ &r_k - n_{\mathrm{H}} d_{\mathrm{H}} \sin(\varphi_k)\cos(\theta_k) - n_{\mathrm{V}} d_{\mathrm{V}} \sin(\theta_k) + \frac{n_{\mathrm{H}}^2 d_{\mathrm{H}}^2 + n_{\mathrm{V}}^2 d_{\mathrm{V}}^2}{2r_k}.
\end{align}
Plugging the approximated distance  \eqref{eq:distance_approx} into the array response vector \eqref{eq:array_response_vector}, the received signal \eqref{eq:received_signal} can be re-written as
\begin{align}
\label{eq:received_signal_matrix_form}
  \vect{y}(t) = \vect{H}\bl{\Phi}\vect{A} \Tilde{\vect{s}}(t) + \vect{w}(t),
\end{align}
where $\vect{A} \in \mathbb{C}^{N \times K}$ is the array response matrix, which collects all the array response vectors corresponding to the $K$ UEs and is given in \eqref{eq:array_response_matrix} at the top of the next page.
\begin{figure*}[t]
    \begin{align}
        \label{eq:array_response_matrix}
 \vect{A} =        \begin{bmatrix}
e^{j\left(- \frac{N_{\mathrm{H}} - 1}{2}\alpha_1 - \frac{N_{\mathrm{V}} -1}{2}\beta_1 - \left(\frac{(N_{\mathrm{H}} -1)^2 d_{\mathrm{H}}^2}{4} + \frac{(N_{\mathrm{V}} -1)^2 d_{\mathrm{V}}^2 }{4}\right)\gamma_1\right)} & \ldots & e^{j\left(- \frac{N_{\mathrm{H}} - 1}{2}\alpha_K - \frac{N_{\mathrm{V}} -1}{2}\beta_K - \left(\frac{(N_{\mathrm{H}} -1)^2 d_{\mathrm{H}}^2}{4} + \frac{(N_{\mathrm{V}} -1)^2 d_{\mathrm{V}}^2 }{4}\right)\gamma_K\right)}  \\
\vdots & \ddots & \vdots  \\
e^{j\left(\frac{N_{\mathrm{H}} - 1}{2}\alpha_1 + \frac{N_{\mathrm{V}} -1}{2}\beta_1 - \left(\frac{(N_{\mathrm{H}} -1)^2 d_{\mathrm{H}}^2}{4} + \frac{(N_{\mathrm{V}} -1)^2 d_{\mathrm{V}}^2 }{4}\right)\gamma_1\right)} & \ldots & e^{j\left(\frac{N_{\mathrm{H}} - 1}{2}\alpha_K + \frac{N_{\mathrm{V}} -1}{2}\beta_K - \left(\frac{(N_{\mathrm{H}} -1)^2 d_{\mathrm{H}}^2}{4} + \frac{(N_{\mathrm{V}} -1)^2 d_{\mathrm{V}}^2 }{4}\right)\gamma_K\right)} 
\end{bmatrix}   
    \end{align}
    \hrulefill
\end{figure*}
In \eqref{eq:array_response_matrix},
 $\alpha_k = \frac{2\pi}{\lambda}d_{\mathrm{H}}\sin(\varphi_k)\cos(\theta_k)$, $\beta_k = \frac{2\pi}{\lambda}d_{\mathrm{V}}\sin(\theta_k)$, and $\gamma_k = \frac{\pi}{\lambda r_k}$. Moreover, $\Tilde{\vect{s}}(t) = [\Tilde{s}_1(t),\ldots,\Tilde{s}_K(t)]^\T$ with $\Tilde{s}_k(t) = \sqrt{p_k \eta_k} s_k(t),\,k = 1,\ldots,K$.

\section{The Proposed Algorithm}
The angle and distance information of the $K$ UEs are hidden in the array response matrix $\vect{A}$. To obtain this information, we need to first extract $\vect{A}\Tilde{\vect{s}}(t)$ from  \eqref{eq:received_signal_matrix_form}. Setting $\vect{G} = \vect{H}\bl{\Phi}$, we can apply the LS method and estimate $\vect{A}\Tilde{\vect{s}}$ as $\overline{\vect{A}\Tilde{\vect{s}}(t)} = \vect{G}^\dagger \vect{y}(t)$, where $\vect{G}^\dagger$ denotes the pseudoinverse of $\vect{G}$. For this estimation to be unique, $\vect{G}$ must have full column rank. However, in a typical RIS-assisted system, we have $M < N$, i.e., the number of BS antennas is smaller than the number of RIS elements. Thus, we have $\mathrm{rank}(\vect{G}) \leq \min(M,N) = M$. In this situation, applying LS to estimate $\vect{A}\Tilde{\vect{s}}$ will not provide us with an accurate estimate because there are fewer observations than unknowns. To overcome this issue, we split each time-slot into $L$ sub-slots, and let the RIS change its configuration between the sub-slots. Therefore, in each time-slot, the RIS takes $L$ different configurations. If we denote  the RIS configuration matrix in the $l$th sub-slot by $\bl{\Phi}_l$, the received signal at the BS in time-slot $t$ will become 
\begin{equation}
    \label{eq:received_signal_over_subslots}
    \Tilde{\vect{y}}(t) = \Tilde{\vect{G}} \vect{A}\Tilde{\vect{s}}(t) + \Tilde{\vect{w}}(t),
\end{equation}where $\Tilde{\vect{G}} = \left(\vect{I}_L \otimes \vect{H}\right)\Tilde{\bl{\Phi}}$ and $\Tilde{\bl{\Phi}} = [\bl{\Phi}_1,\ldots,\bl{\Phi}_L]^\T$.
If $L$ is sufficiently large such that $\Tilde{\vect{G}}$ is a full column rank matrix, the LS method can be applied to uniquely estimate $\vect{A}\Tilde{\vect{s}}(t)$ as 
\begin{equation}
\label{eq:estimate}
    \overline{\vect{A}\Tilde{\vect{s}}(t)} = \Tilde{\vect{G}}^\dagger \Tilde{\vect{y}}(t)  = (\Tilde{\vect{G}}^\H \Tilde{\vect{G}})^{-1} \Tilde{\vect{G}}^\H \Tilde{\vect{y}}(t). 
\end{equation}
From \eqref{eq:received_signal_over_subslots} and \eqref{eq:estimate}, the covariance matrix of the signal incident on the RIS can be obtained as 
\begin{equation}
\label{eq:covariance_matrix}
   \vect{R}  = \mathbb{E}\left\{ \overline{\vect{A}\Tilde{\vect{s}}(t)} \left(\overline{\vect{A}\Tilde{\vect{s}}(t)}\right)^\H \right\} = \vect{ASA}^\H + \sigma^2 \left(\Tilde{\vect{G}}^\H \Tilde{\vect{G}}\right)^{-1},
\end{equation}where $\vect{S} = \mathbb{E}\left\{ \Tilde{\vect{s}}(t) \Tilde{\vect{s}}(t)^\H \right\}$. The eigendecomposition of $\vect{R}$ yields
\begin{equation}
\label{eq:main_covariance_matrix}
  \vect{R} = \vect{U}_s \bl{\Sigma}_s \vect{U}_s^\H + \vect{U}_n \bl{\Sigma}_n \vect{U}_n^\H,  
\end{equation}
where $\bl{\Sigma}_s \in \mathbb{C}^{K \times K}$  and $\bl{\Sigma}_n \in \mathbb{C}^{(N-K) \times (N-K)}$ are diagonal matrices having the $K$ largest and $(N- K)$ smallest eigenvalues of $\vect{R}$ on their diagonal, respectively. The matrices $\vect{U}_s \in \mathbb{C}^{N \times K}$ and $\vect{U}_n \in \mathbb{C}^{N \times (N-K)}$ contain the eigenvectors corresponding to the eigenvalues on the diagonal of $\bl{\Sigma}_s$ and $\bl{\Sigma}_n$, respectively. 

In practice, we need to estimate the covariance matrix by averaging over $T$ samples. The sample covariance matrix is given by 
\begin{equation}
\label{eq:sample_covariance_matrix}
    \hat{\vect{R}} = \frac{1}{T} \sum_{t = 1}^T \overline{\vect{A}\Tilde{\vect{s}}(t)} \left(\overline{\vect{A}\Tilde{\vect{s}}(t)}\right)^\H.
\end{equation}

The conventional MUSIC algorithm can be used to estimate the AoAs and distances by utilizing the orthogonality between signal and noise subspaces of the covariance matrix. This algorithm  performs a 3D search over all the possible values of the parameters to obtain $K$ tuples $(\hat{\varphi}_k,\hat{\theta}_k,\hat{r}_k),\,k = 1,\ldots,K$, where $\hat{x}$ denotes the estimated value of $x$. However, the 3D search incurs a tremendously high computational complexity since the number of grid points grows cubically with the resolution per dimension. 

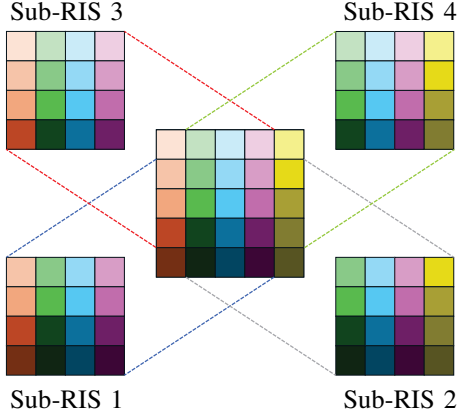
\begin{figure}[t!]
	\centering
	\begin{overpic}[width=0.75\columnwidth]{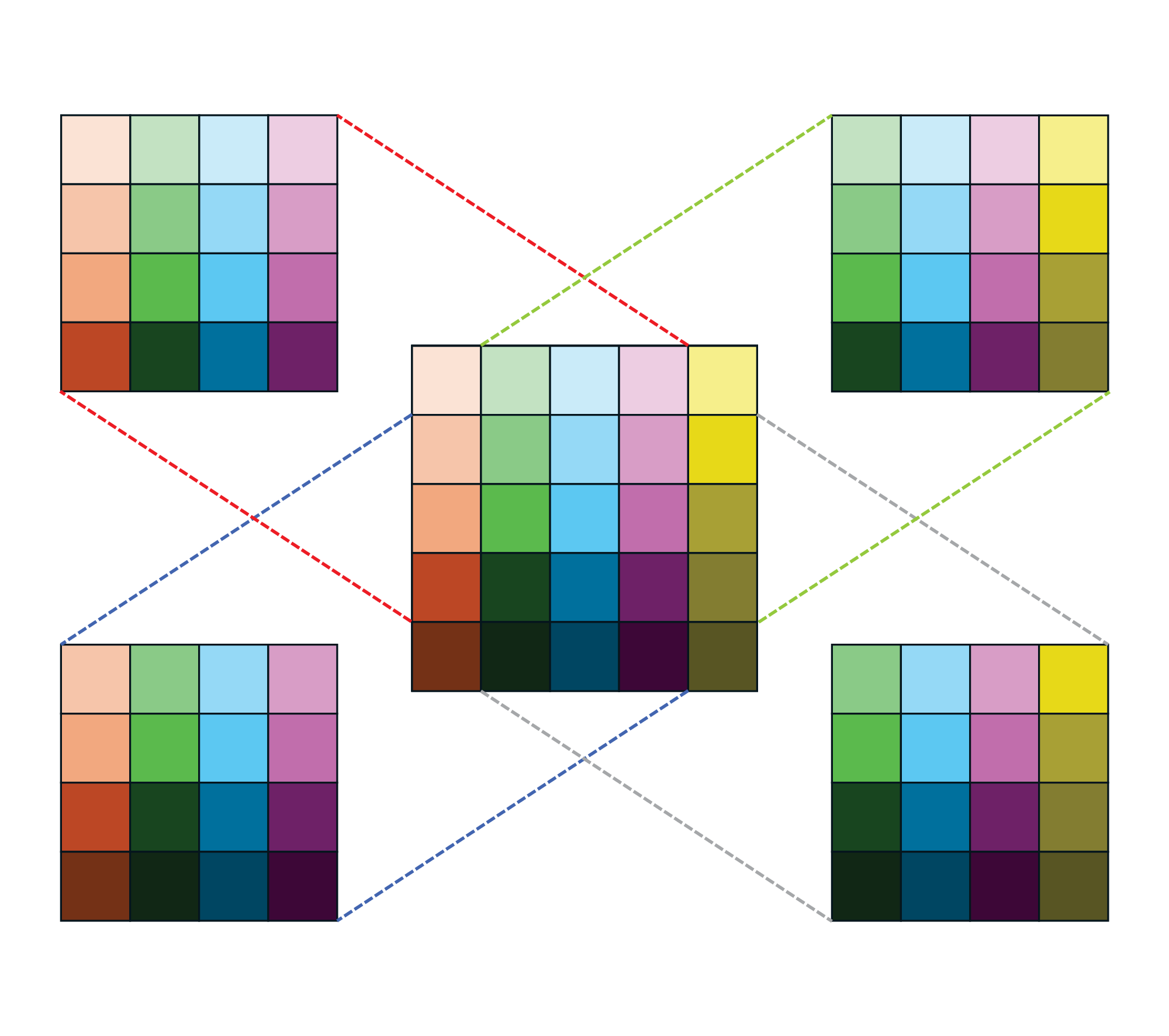}
  \put (6,3.4) {Sub-RIS 1}
   \put (72.7,3.4) {Sub-RIS 2}
   \put (6,81) {Sub-RIS 3}
   \put (72.7,81) {Sub-RIS 4}
    \end{overpic}
    \caption{To perform spatial smoothing, we divide an RIS into overlapping sub-RISs. In this example, a $5 \times 5$ RIS is split into four sub-RISs of size $4 \times 4$.    }
    \label{fig:spatial_smoothing}
\end{figure}

Herein, we propose a modified MUSIC algorithm that utilizes the symmetric structure of the array response vector $\vect{A}$ to decouple AoA and distance estimation problems. Disregarding the noise term as in \cite{He2012efficient}, the anti-diagonal entries of the covariance matrix $\vect{R}$ in \eqref{eq:covariance_matrix} are given by  
\begin{align}
\label{eq:anti_diagonal_entries}
    &[\vect{R}]_{n,N+1-n}= \mathbb{E}\Bigg\{ \left( \sum_{k=1}^K e^{j\left(n_{\mathrm{H}} \alpha_k + n_{\mathrm{V}} \beta_k - (n_{\mathrm{H}}^2 + n_{\mathrm{V}}^2)\gamma_k\right)}\tilde{s}_k\right)  \nonumber\\& \times \left( \sum_{k=1}^K e^{-j\left(-n_{\mathrm{H}} \alpha_k - n_{\mathrm{V}} \beta_k - (n_{\mathrm{H}}^2 + n_{\mathrm{V}}^2)\gamma_k\right)}\tilde{s}_k^*\right) \Bigg\}.
\end{align}
Noting the independence between the different signals, the expression in \eqref{eq:anti_diagonal_entries} is simplified as 
\begin{equation}
   [\vect{R}]_{n,N+1-n} = \sum_{k=1}^K q_k e^{j2(n_{\mathrm{H}}\alpha_k + n_{\mathrm{V}}\beta_k )}, 
\end{equation}with $q_k  = \mathbb{E}\{\Tilde{s}_k \Tilde{s}_k^*\} = p_k \eta_k$. Note that $n_{\mathrm{H}}$ and $n_{\mathrm{V}}$ are the horizontal and vertical indices of the $n$th RIS element, given by $n_{\mathrm{H}} = \mod(n-1,N_{\mathrm{H}}) - \frac{N_\mathrm{H} - 1}{2} $, $n_{\mathrm{V}} = \left \lfloor \frac{n-1}{N_{\mathrm{H}}} \right \rfloor - \frac{N_V - 1}{2}$.
 Starting from $n = 1$, we collect all the $N$ anti-diagonal entries of $\vect{R}$ in a vector $\bar{\vect{y}}$ as  
\begin{align}
\label{eq:stacked_signals}
  \bar{\vect{y}} = \Big[ &\sum_{k=1}^K q_k e^{j2\left(- \frac{N_{\mathrm{H}} - 1}{2}\alpha_k - \frac{N_{\mathrm{V}} - 1}{2}\beta_k\right)}, \ldots, \nonumber \\ &\sum_{k=1}^K q_k e^{j2\left( \frac{N_{\mathrm{H}} - 1}{2}\alpha_k + \frac{N_{\mathrm{V}} - 1}{2}\beta_k\right)} \Big]^\T.  
\end{align}
We can see that the entries of $\bar{\vect{y}}$ only depend on the azimuth and elevation AoAs and are independent of the distances between the UEs and the RIS. Therefore, we can decouple the AoA and distance estimation problems. The covariance matrix of the signal vector in \eqref{eq:stacked_signals} is a rank-one matrix and cannot be used for estimating the $K$ AoAs. 
Hence, we need a spatial smoothing procedure to improve the rank of the sample covariance matrix. To this end, we divide the RIS into sub-RISs and extract the corresponding sub-vector for each sub-RIS from the vector $\bar{\vect{y}}$. Specifically, we first split the RIS into $J$ overlapping sub-RISs, each having the size of $D_{\mathrm{H}} \times D_{\mathrm{V}}$, i.e., having $D_{\mathrm{H}}$ elements in each row and $D_{\mathrm{V}}$ elements in each column. In an RIS of size $N_{\mathrm{H}} \times N_{\mathrm{V}}$, we have $J = (N_{\mathrm{H}} - D_{\mathrm{H}} +1) \times (N_{\mathrm{V}}-D_{\mathrm{V}}+1)$ overlapping sub-RISs of size $D_{\mathrm{H}} \times D_{\mathrm{V}}$. Fig.~\ref{fig:spatial_smoothing} presents an illustrative example where an RIS with $N_{\mathrm{H}} = N_{\mathrm{V}} = 5$ is split into $J = 4$ overlapping sub-RISs with $D_{\mathrm{H}} = D_{\mathrm{V}} = 4$. 

For each sub-RIS, we extract the corresponding entries from the vector $\bar{\vect{y}}$ and create a new sub-vector of length $D_{\mathrm{H}} \times D_{\mathrm{V}}$. For the $i$th sub-RIS, the corresponding sub-vector is given by 
   \begin{align}
    \label{eq:sub-vector}
&\bar{\vect{y}}_i = \bigg[\sum_{k=1}^K q_k e^{j2(n_{\mathrm{H},i}\alpha_k + n_{\mathrm{V},i}\beta_k)}, \nonumber \\ &\ldots,\sum_{k=1}^K q_k e^{j2\big((n_{\mathrm{H},i}+D_{\mathrm{H}}-1)\alpha_k + n_{\mathrm{V},i}\beta_k \big)},\nonumber \\ &\ldots,\sum_{k=1}^K q_k e^{j2\big((n_{\mathrm{H},i}+D_{\mathrm{H}} - 1)\alpha_k + (n_{\mathrm{V},i} + D_{\mathrm{V}} - 1)\beta_k \big)}\bigg]^\T,
    \end{align}
where $n_{\mathrm{H},i}$ and $n_{\mathrm{V},i}$ represent the horizontal and vertical indices of the first element of the $i$th sub-RIS in the original RIS. For example, in Fig.~\ref{fig:spatial_smoothing}, we have $n_{\mathrm{H},1} = -2,\,n_{\mathrm{V},1} = -2$, $n_{\mathrm{H},2} = -1,\,n_{\mathrm{V},2} = -2$, and so on. The sub-vector $\bar{\vect{y}}_i$ can be decomposed as 
\begin{equation}
   \bar{\vect{y}}_i = \vect{B}\vect{q}_i,   
\end{equation}where 
\begin{align}
 \vect{B} &= [\vect{b}(\alpha_1,\beta_1),\ldots,\vect{b}(\alpha_K,\beta_K)] \in \mathbb{C}^{D_{\mathrm{H}} D_{\mathrm{V}} \times K }, \\
 \vect{q}_i &= [q_1 e^{j2(n_{\mathrm{H},i}\alpha_1 + n_{\mathrm{V},i}\beta_1)},\ldots,q_K e^{j2(n_{\mathrm{H},i}\alpha_K + n_{\mathrm{V},i}\beta_K)}]^\T \nonumber \\ &\in \mathbb{C}^{K},
\end{align}and 
\begin{align}
  &\vect{b}(\alpha_k,\beta_k) = \nonumber \\
  &\left[1,\ldots,e^{j2(D_{\mathrm{H}} - 1)\alpha_k},\ldots,e^{j2\big((D_{\mathrm{H}} - 1)\alpha_k + (D_{\mathrm{V}} - 1)\beta_k\big)} \right]^\T.  
\end{align} The sample covariance matrix of the sub-vectors can be expressed as 
\begin{align}
    &\bar{\vect{R}} = \frac{1}{J}\sum_{i=1}^{J}  \bar{\vect{y}}_i \bar{\vect{y}}_i^\H =\frac{1}{J}\vect{B}\underbrace{\left(\sum_{i=1}^{J}\vect{q}_i \vect{q}_i^\H \right)}_{\triangleq \vect{R}_q}\vect{B}^\H  = \frac{1}{J}\vect{B}\vect{R}_q\vect{B}^\H, 
\end{align}
and the eigendecomposition of $\bar{\vect{R}}$ yields 
\begin{equation}
    \bar{\vect{R}}  = \bar{\vect{U}}_s \bar{\bl{\Sigma}}_s \bar{\vect{U}}_s^\H + \bar{\vect{U}}_n \bar{\bl{\Sigma}}_n \bar{\vect{U}}_n^\H,  
\end{equation}
 where $\bar{\bl{\Sigma}}_s \in \mathbb{C}^{K \times K}$  and $\bar{\bl{\Sigma}}_n \in \mathbb{C}^{(D_{\mathrm{H}}D_{\mathrm{V}}-K) \times (D_{\mathrm{H}}D_{\mathrm{V}}-K)}$ contain the $K$ largest and $(D_{\mathrm{H}}D_{\mathrm{V}}-K)$ smallest eigenvalues of $\bar{\vect{R}}$ on their diagonal, and $\bar{\vect{U}}_s \in \mathbb{C}^{D_{\mathrm{H}}D_{\mathrm{V}} \times K}$ and $\bar{\vect{U}}_n \in \mathbb{C}^{D_{\mathrm{H}}D_{\mathrm{V}} \times (D_{\mathrm{H}}D_{\mathrm{V}}-K)}$ consist of the corresponding eigenvectors. We can now employ the MUSIC algorithm to estimate the $K$ azimuth and elevation AoAs. Specifically, the orthogonality between signal and noise subspaces results in
\begin{align}
    &\vect{B}^\H \bar{\vect{U}}_n = \vect{0} \nonumber \\
    &\Rightarrow \quad \vect{b}^\H(\alpha_k,\beta_k)\bar{\vect{U}}_n \bar{\vect{U}}_n^\H \vect{b}(\alpha_k,\beta_k) = 0, \quad k = 1,\ldots, K.
\end{align}
Hence, the AoAs can be found by identifying the $K$ highest peaks of the spectrum
\begin{equation}
\label{eq:modified_MUSIC_spectrum}
 f(\varphi,\theta) =  \frac{1}{\vect{b}^\H(\alpha,\beta) \bar{\vect{U}}_n \bar{\vect{U}}_n^\H \vect{b}(\alpha,\beta) }. 
\end{equation}
\vspace{-0.5cm}
\begin{figure}
\centering
\label{fig:angle}\subfloat[Angular peaks]
{\includegraphics[scale=0.49]{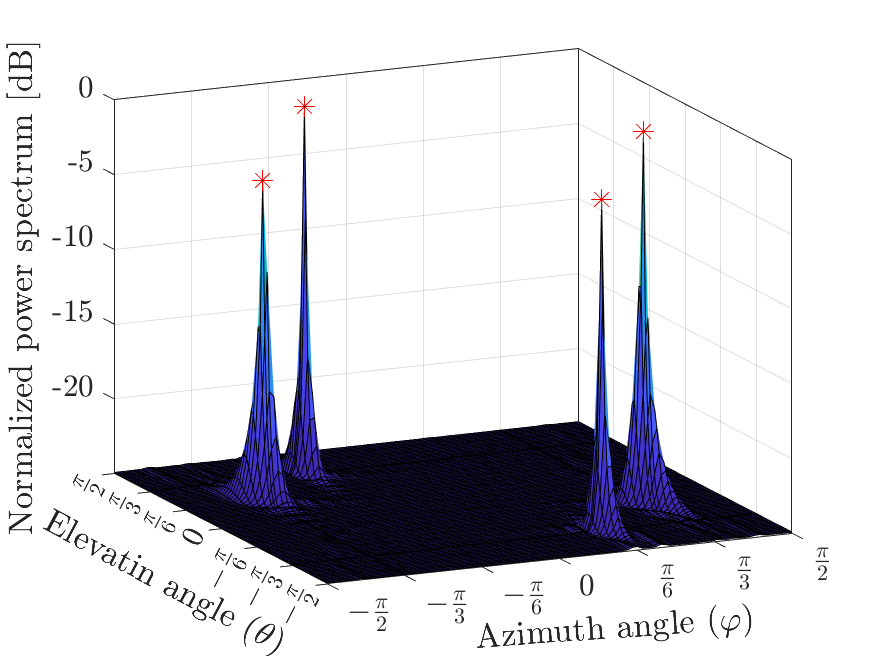}}\hfill \\
\centering
\label{fig:distance}\subfloat[Distance peaks]
{\includegraphics[scale=0.49]{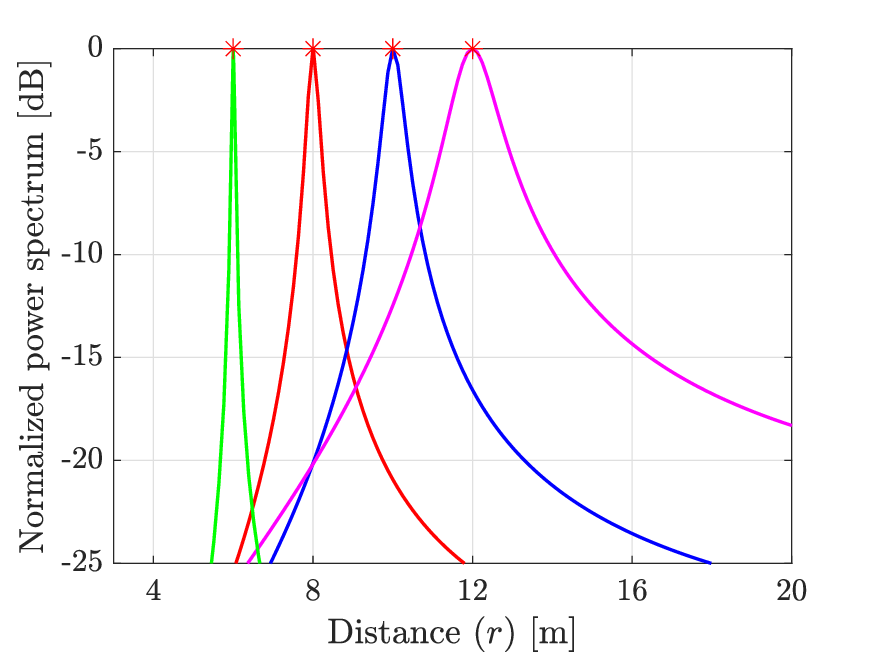}}
\caption{The modified MUSIC spectrum for $N_{\rm H} = N_{\rm V} = 25, M=128, D_{\mathrm{H}} = D_{\mathrm{V}} = 22$. Red stars show the exact location of the UEs.}
\label{fig:Peaks}
\end{figure}
\begin{remark}
    To identify the $K$ AoAs using the presented modified MUSIC algorithm, we need to have $\mathrm{rank}(\vect{B}) = K$ and $\mathrm{rank}(\vect{R}_q) = K$. The former is satisfied when $D_{\mathrm{H}} D_{\mathrm{V}} > K$. Furthermore, $\vect{R}_q$ is a summation of $J$ rank-one matrices where $J = (N_{\mathrm{H}} - D_{\mathrm{H}} +1)(N_{\mathrm{V}} - D_{\mathrm{V}} + 1)$. Thus, it becomes a full-rank matrix when $(N_{\mathrm{H}} - D_{\mathrm{H}} +1)(N_{\mathrm{V}} - D_{\mathrm{V}} + 1) > K$.
\end{remark}

After estimating the azimuth and elevation AoAs, we can employ the standard MUSIC algorithm to find the distances corresponding to the estimated AoAs. In particular, for the AoA estimate $(\hat{\varphi}_k,\hat{\theta}_k)$, the respective distance is obtained as 
\begin{equation}
\label{eq:distance_estimation}
    \hat{r}_k = \argmax{r}\,\frac{1}{\vect{a}^\H(\hat{\varphi}_k,\hat{\theta}_k,r)\hat{\vect{U}}_n \hat{\vect{U}}_n^\H \vect{a}(\hat{\varphi}_k,\hat{\theta}_k,r)}, 
\end{equation}where the columns of $\hat{\vect{U}}_n \in \mathbb{C}^{N \times (N - K)}$ are the eigenvectors corresponding to the $N - K$ smallest eigenvalues of the sample covariance matrix $\hat{\vect{R}}$ given in \eqref{eq:sample_covariance_matrix}.

\section{Numerical Results}

In this section, we numerically evaluate the performance of the proposed modified MUSIC algorithm. In all the simulations, we use  $d_{\rm H} = d_{\rm V} =0.5 \lambda$,  $L= \left \lceil \frac{N}{M} \right \rceil$, and the $L$ RIS configurations are selected from the columns of the discrete Fourier transform (DFT) matrix. We adopt the free-space path-loss model for the channels between the RIS and the UEs; thus, we have $\eta_k = \frac{\lambda^2}{(4\pi r_k)^2}$, where $\lambda = 0.3\,$m. The channel between the RIS and the BS follows Rician fading with the Rician factor of $2$.  The number of samples is $T = 300$.
The noise power is $\sigma^2 = -154\,$dBm, and unless otherwise stated, the transmit power of the UEs is set to be $p_k = 10\,$dBm,~$\forall k$.

Fig.~\ref{fig:Peaks} depicts the angle and distance spectra of the proposed modified MUSIC algorithm. 
The following setup is used for the RIS and the BS: 
$N_\mathrm{H} = N_{\mathrm{V}} = 25$, $M = 128$, $D_{\mathrm{H}} = D_{\mathrm{V}} = 22$. There are $K = 4$ UEs located at $(\pi/6,-\pi/3,6\,\mathrm{m}), (-\pi/6,\pi/3,8\,\mathrm{m}), (\pi/3,-\pi/6,10\,\mathrm{m}),$ and $(-\pi/3,\pi/6,12\,\mathrm{m})$, where the tuple entries indicate azimuth AoA, elevation AoA, and distance, respectively. Fig.~\ref{fig:Peaks}(a) shows that the modified MUSIC accurately resolves the AoAs of the UEs. Since the AoAs are successfully estimated 
in the first step, the corresponding distances can be accurately estimated using \eqref{eq:distance_estimation}, as demonstrated in Fig.~\ref{fig:Peaks}(b).

We now compare the performance of the proposed scheme with the standard 3D MUSIC, which has a significantly higher complexity due to searching across the 3D spatial domain. Specifically, with the same number of grid points in the azimuth, elevation, and distance domains, our simulations showed that the runtime of the proposed modified MUSIC is $184$ times less than that of 3D MUSIC.
\begin{figure}[!t]
    \centering
    \includegraphics[width=0.85\columnwidth]{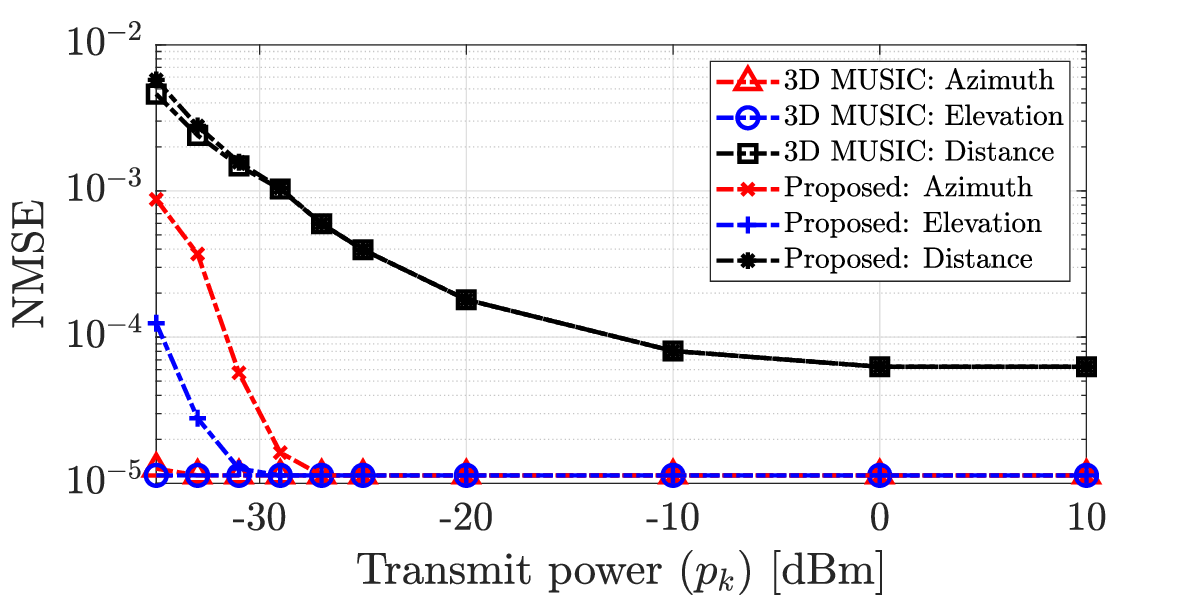}
    \caption{The NMSE versus transmit power with $N_{\rm H} = N_{\rm V} = 7,\, M=16, \,D_{\mathrm{H}} = D_{\mathrm{V}} = 6,$ and $K=2$.}
    \label{fig:F_3DvsMod}
\end{figure}
\begin{figure}[!t]
    \centering
    \includegraphics[width=0.85\columnwidth]{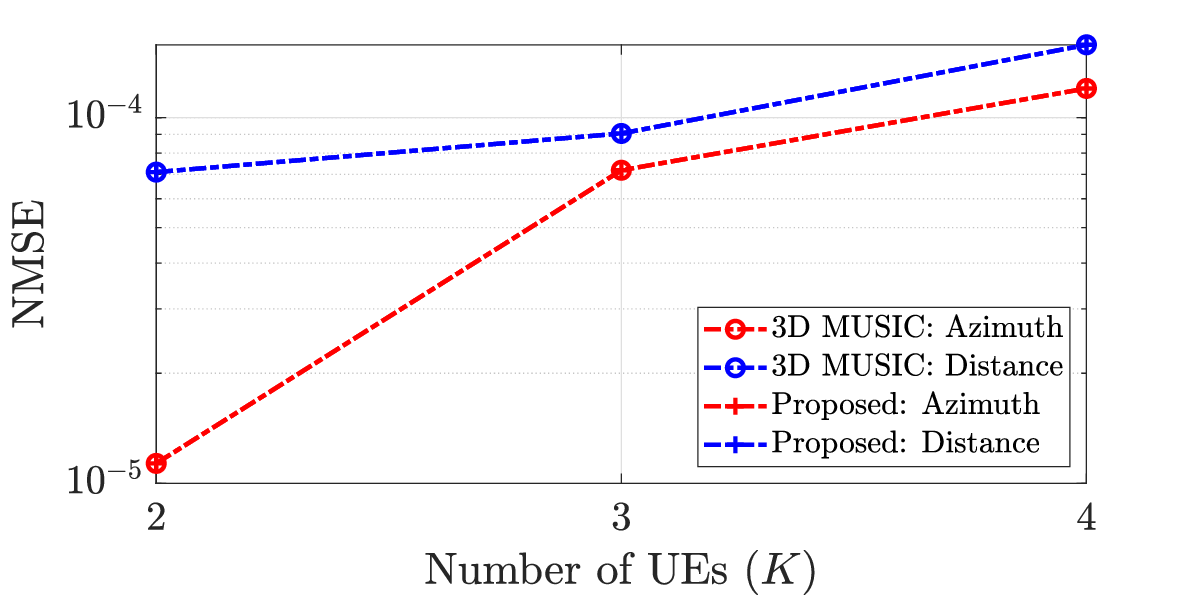}
    \caption{The NMSE versus number of UEs with $N_{\rm H} = N_{\rm V} = 25,\,M=128,\, D_{\mathrm{H}} = D_{\mathrm{V}} = 22$.}
    \label{fig:F_2}
\end{figure}
 Fig.~\ref{fig:F_3DvsMod} depicts the normalized mean-square-error (NMSE) of the proposed method and the 3D MUSIC algorithm. The NMSE is averaged over $10000$ channel and noise realizations.
A \(7\times7\) RIS is considered and the number of BS antennas is assumed to be \(M=16\). For spatial smoothing, we divide the RIS into four $6 \times 6$ sub-RISs. 
 We consider two UEs positioned at \((\pi/6, -\pi/6, 1.5\,\mathrm{m})\) and \((-\pi/6, \pi/6, 2\,\mathrm{m})\).
We observe that when the transmit power is sufficiently high (e.g., above $-25$\,dBm), the NMSE of the modified MUSIC scheme closely matches that of the standard 3D MUSIC algorithm. This demonstrates the effectiveness of the proposed modified MUSIC, as we can achieve almost identical NMSE performance to the 3D MUSIC with significantly reduced complexity.

We now investigate the performance of the proposed modified MUSIC algorithm versus the number of UEs. In this simulation, we use the same parameters as in Fig.~\ref{fig:Peaks}. 
In such a large-dimensional system, running the 3D MUSIC is infeasible due to the excessively high complexity of the 3D search. Therefore, we assume that the elevation AoA is known, thereby simplifying the 3D MUSIC to 2D MUSIC and searching for spectrum peaks over the azimuth AoA and distance domains. 
The UEs are located at \((\pi/6, 0, 6\,\mathrm{m})\), \((-\pi/6, 0, 6.5\,\mathrm{m})\), \((\pi/12, 0, 7\,\mathrm{m})\), and \((-\pi/12, 0, 7.5\,\mathrm{m})\). When $K <4$, only the first \(K\) UEs are considered for localization. The results are illustrated in Fig.~\ref{fig:F_2}. It is evident that the modified MUSIC achieves the same NMSE performance as the standard 3D MUSIC algorithm, in both angle and distance domains.

\begin{figure}
\centering
\subfloat[ $N_{\rm H} = N_{\rm V} = 15,\, D_{\mathrm{H}} = D_{\mathrm{V}} = 12$.]
{\includegraphics[scale=0.52]{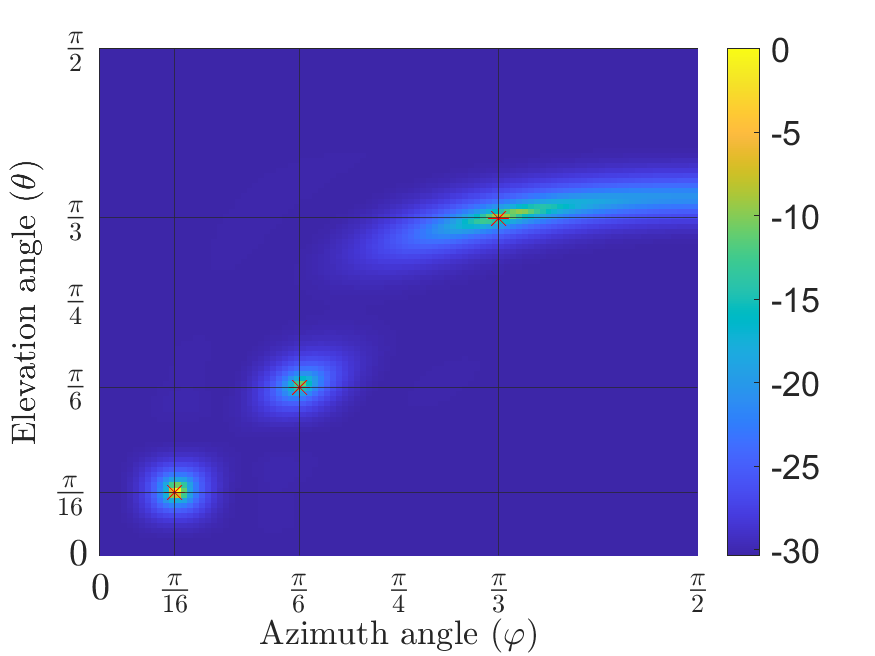}}\hfill
\centering \\
\subfloat[ $N_{\rm H} = N_{\rm V} = 35,\, D_{\mathrm{H}} = D_{\mathrm{V}} = 32$.]
{\includegraphics[scale=0.52]{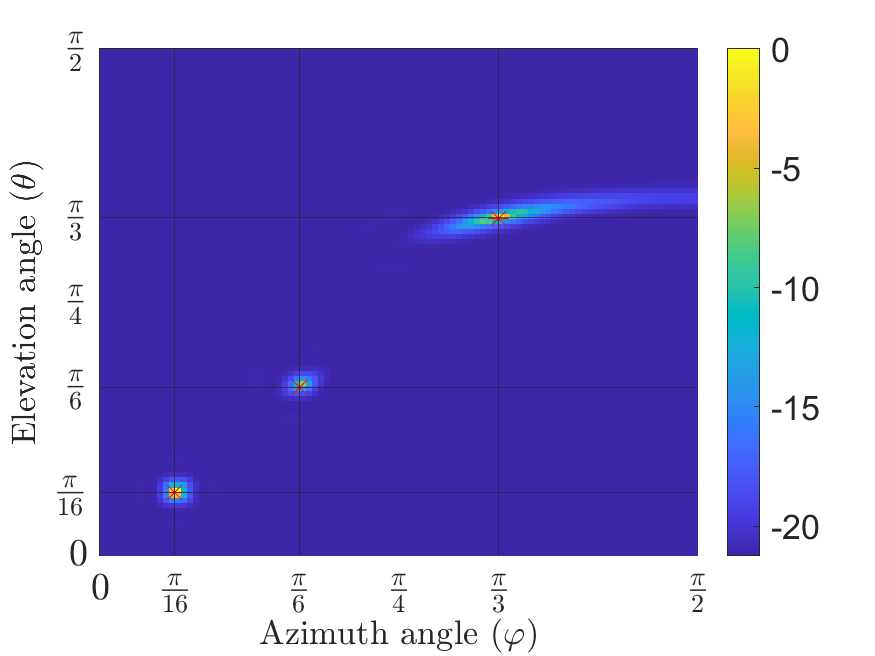}}
\caption{The angular spectrum (dB) in \eqref{eq:modified_MUSIC_spectrum} for three UEs located at $(\pi/16,\pi/16)$, $(\pi/6,\pi/6)$, and $(\pi/3,\pi/3)$. Red stars show the exact location of the UEs. }
\label{F_3}
\end{figure}

An important observation is that when a UE moves towards the end-fire direction of the RIS, the energy of the spectrum in \eqref{eq:modified_MUSIC_spectrum} at the UE location becomes more dispersive. Fig.~\ref{F_3} shows the angular spectrum in \eqref{eq:modified_MUSIC_spectrum} when there are three UEs located in the angular directions $(\pi/16,\pi/16)$, $(\pi/6,\pi/6)$, and $(\pi/3,\pi/3)$. The number of BS antennas is set to be $M = 128$. We first assume that the RIS is equipped with $N_\mathrm{H} = N_{\mathrm{V}} = 15$ elements and is divided into $J = 16$ sub-RISs with $D_{\mathrm{H}} = D_{\mathrm{V}} = 12$. Fig.~\ref{F_3}(a) shows that the detected peak for the UE at $(\pi/16,\pi/16)$ accurately matches the angular direction of the UE. 
However, the spectrum energy is less focused for the UE that is located at $(\pi/3,\pi/3)$, resulting in a lower location estimation accuracy for this UE.  
 This is aligned with the  finite beam depth analysis in \cite{2024_Kosasih_TWC}, where it was shown that the minimum beam depth is achieved when the transmitter is in the broadside of the antenna array. We can improve the estimation accuracy at the expense of increased complexity by increasing the number of RIS elements. This is illustrated in Fig.~\ref{F_3}(b), where we have $1225$ RIS elements, and $J = 16$ sub-RISs, each having $1024$ elements. With more RIS elements, the spectrum energy becomes more focused, and the locations of the UEs are more accurately estimated. 

\section{Conclusions}
This paper considered an RIS-assisted near-field localization system and proposed a novel subspace-based localization scheme that, by exploiting the symmetric structure of the RIS, remarkably reduces the complexity compared to the standard 3D MUSIC algorithm. In particular, we first applied the LS method to obtain the incident signal on the RIS and then, decoupled the AoA and distance estimation problems using the structure of the RIS array response matrix. We further proposed a spatial smoothing scheme to address the rank-deficiency of the signal covariance matrix and efficiently estimate the AoAs of multiple UEs. We demonstrated via numerical simulations that our scheme has a comparable NMSE performance to the standard 3D MUSIC algorithm, though incurring much lower complexity.

\bibliographystyle{IEEEtran}

\bibliography{IEEEabrv,bib}

\end{document}